# Fuels of the Future for Renewable Energy Sources (Ammonia, Biofuels, Hydrogen)

**G. Ali Mansoori** [*], University of Illinois at Chicago, Chicago, IL, USA. mansoori@uic.edu

**L. Barnie Agyarko,** City Colleges of Chicago, Chicago, IL, USA.

**L. Antonio Estevez**, University of Puerto Rico, Mayagüez, Puerto Rico, USA.

**Behrooz Fallahi,** Northern Illinois University, DeKalb, IL, USA.

**Georgi Gladyshev,** Russian Academy of Sciences | N. N. Semenov Institute of Chemical Physics, Moscow, Russia.

**Ronaldo Gonçalves dos Santos,** University Center of FEI, São Bernardo do Campo, São Paulo, Brazil.

**Shawn Niaki**, Senior Principal / Managing Director of ECE Project Management G.m.b.H. & Co. KG., Tbilisi, Georgia.

**Ognjen Perišić**, Big Blue Genomics, Vojvode Brane 32, 11000 Belgrade, Serbia,

**Mika Sillanpää,** Duy Tan University, Da Nang, Vietnam; University of Southern Queensland, Toowoomba, Australia; University of Johannesburg, Doornfontein, South Africa.

**Kaniki Tumba,** Mangosuthu University of Technology, Durban, South Africa.

**Jeffrey Yen,** Total Consulting, Shanghai, China.

_______________________

(*). Correspondence: mansoori@uic.edu & gali.mansoori@gmail.com

**Abstract:** Potential strategies for the development and large-scale application of renewable energy sources aimed at reducing the usage of carbon-based fossil fuels are assessed here, especially in the event of the abandonment of such fuels. The aim is to aid the initiative to reduce the harmful effects of carbon-based fossil fuels on the environment and ensure a reduction in greenhouse gases and sustainability of natural resources. Small-scale renewable energy application for heating, cooling, and electricity generation in households and commercial buildings are already underway around the world. Hydrogen ($H_2$) and ammonia ($NH_3$), which are presently produced using fossil fuels, already have significant applications in society and industry, and are therefore good candidates for large-scale production through the use of renewable energy sources. This will help to reduce the greenhouse gas emissions appreciably around the world. While the first-generation biofuels production using food crops may not be suitable for long-range fuel production, due to competition with the food supply, the $2^{nd}$, $3^{rd}$ and $4^{th}$ generation biofuels have the potential to produce large, worldwide supplies of fuels. Production of advanced biofuel**s** will not increase the emission of greenhouse gases, and the ammonia produced through the use of renewable energy resources will serve as fertilizer for biofuels production. The perspective of renewable energy sources, such as technology status, economics, overall environmental benefits, obstacles for commercialization, relative competitiveness of various renewable energy sources, etc., are also discussed whenever applicable.

# 1. Introduction

At present the world is relying mainly on fossil fuels and partly on other sources, such as nuclear fission energy, hydro-power, solar, windmill, etc., to run the industry, transportation, households and commercial lighting, heating and cooling systems. It is now obvious that due to drastic environmental concerns that the use of fossil fuels has been causing, such as excessive addition of greenhouse gases to the atmosphere and the resultant global climate change, the world needs to go towards the use of renewable energies. The 2015 United Nations Climate Change Conference (COP21 Paris Agreement) provided fundamental guidelines to achieve energy sufficiency, clean environment, and sustainable sources for countries, states, or communities. Various options available to achieve goals of energy sufficiency, and clean environment, were presented in the COP21 Paris Agreement. Schemes related to each energy source and their conversion technologies were presented and sustainability of renewable energy sources was discussed. All the possible energy sources including coal, natural gas, petroleum, nuclear, solar, wind, biofuels, and geothermal energy were considered, as well as energy storage options. Also presented were ways of dealing with greenhouse gases, which are produced mostly from fossil fuels combustion, including their collection, transportation, storage, and sequestration. A comprehensive book covering the subject of energy sources, utilization, legislation, sustainability, and Illinois as model state was published a year after the COP21 Paris Agreement, which represents a roadmap to follow its requirements [1].

Presently, the driving force moving the industry in terms of production of materials and chemicals of major need in society is strongly based on non-renewable sources [2]. The fulfillment of the COP21 Paris Agreement requires the reduction and eventual end to dependence on fossil fuels, which will be necessary to decrease the harmful greenhouse gases emission that causes climate change. In order to reach the COP21 requirements it is necessary to develop successful technologies for direct utilization of renewable energies, including solar, wind, geothermal, ocean energy contents, and biomass-derived sources [1, 3-5].

Two of the chemicals which are envisioned to be readily producible using renewable energies are hydrogen ($H_2$) and ammonia ($NH_3$), which are independent of carbon materials and have widespread applications in industry and society [6].

Expansion in the production of biofuels will be a great help towards the use of renewable energies through agricultural activities [7]. Biofuels are derived from plants, which are simply recyclers of carbon dioxide in the air using sunlight and they could readily replace fossil fuels [3, 8].

Energy conservation and improvements in energy consumption efficiencies [9] are always necessary for better use of any energy source including renewable energies. There are, at least, three major approaches to use renewable energies: Small-scale heating, cooling, electricity generation systems to serve the residential commercial and transportation industries of individual entities and small communities. Hydrogen and ammonia production using renewable energies to replace the existing fossil-fuels based technologies. Expansion in agriculture to produce the plants needed for advanced biofuels industry. The promising technologies to achieve these three main goals will be presented in the following sections.

## 2. Small-Scale Renewable Energy Utilization

Small-scale use of renewable energies for heating, cooling and lighting of buildings presently represents a large fraction of the total renewable energy applications. Geothermal, seasonal energy storage, solar, and wind energies are typical renewable energy source that have been developed to small-scale usage (Figure 1).

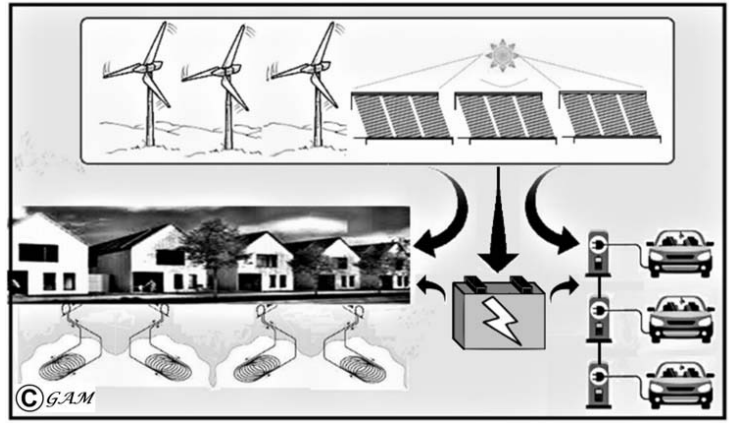

**Figure 1:** Small-scale utilization of renewable energies (geothermal, seasonal energy storage, solar, wind) is now in practice around the world.

It is worth mentioning that the technology for heating and cooling using geothermal energy is rather well developed and in widespread use. The technology for heating through solar energy is also well understood and many buildings are now equipped with solar collectors for heating and electricity-generating purposes. Cooling and air conditioning using solar absorption cycle technology was advanced during 1970s and 1980s [1,10,11]. Wind energy utilization is especially useful for electricity generation. Overall, the technologies for all such methods of renewable energy utilization are now available and many communities around the world are becoming energy-independent using such approaches.

## 3. Large-Scale Renewable Energy Utilization

The large-scale use of the renewable energies in transportation, industrial and commercial applications requires appropriate fuel(s) obtained on specific standards. To be used in these applications the fuel needs to be independent from the fossil fuels usage. For these reasons there are speculations to use hydrogen, ammonia and $2^{nd}$, $3^{rd}$ & $4^{th}$ generation biofuels, which are based on the use of agricultural waste and certain energy-crops.

### 3.1. Hydrogen ($H_2$) as a fuel of the future:

Hydrogen is quite suitable as the fuel for fuel-cells and internal combustion engines. It can be readily produced using renewable energy sources through dissociation of water [12] or gasification process [5]. However, there are major difficulties for transportation and storage of hydrogen with chemical formula and molecular structure given in Figure 2.

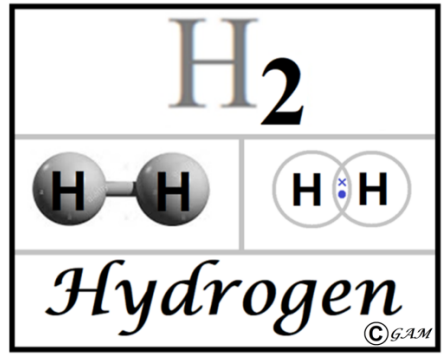

**Figure 2:** Various styles of the molecular and chemical structure of hydrogen molecule

Hydrogen is a very light molecule, odorless and vulnerable to leakage. Conceivably, upon its production, hydrogen is compressed up to about 3-8 MPa pressure, and then fed into a transmission pipeline. The pipeline transports the hydrogen to a compressed gas terminal where the hydrogen is loaded into compressed gas tube trailers. Alternately, trucks can deliver the tube trailers to a forecourt station where the hydrogen is further compressed, stored, and dispensed to fuel-cell vehicles. In certain part so Europe, some gas stations offer electric automotive charging as well as hydrogen refuels.

Based on existing technologies, the production cost of green hydrogen via solar or windmill-powered electrolysis is approximately 2 to 3 times that from steam methane reforming (SMR) while the cost of the green hydrogen from landfill gas dry fermentation is close to that from SMR [13]. If the hydrogen energy can be used widely, the Hydrogen Council (hydrogencouncil.com) estimates that the costs to produce and distribute hydrogen from clean energy sources will fall by as much as 50% over the next decade. Hydrogen need to be efficiently stored, especially when the energy is derived from the solar system which is only available during the day, before it is transferred into a fuel-cell, which then uses the hydrogen to create clean electricity with no emissions.

The long-term hydrogen storage pathway focuses on both (1) cold or cryo-compressed hydrogen storage, where increased hydrogen density and insulated pressure vessels may allow for US-DOE targets to be met and (2) materials-based hydrogen storage technologies, including sorbents, chemical hydrogen storage materials, and metal hydrides, with properties having the potential to meet the US-DOE hydrogen storage targets as shown in Figure 3.

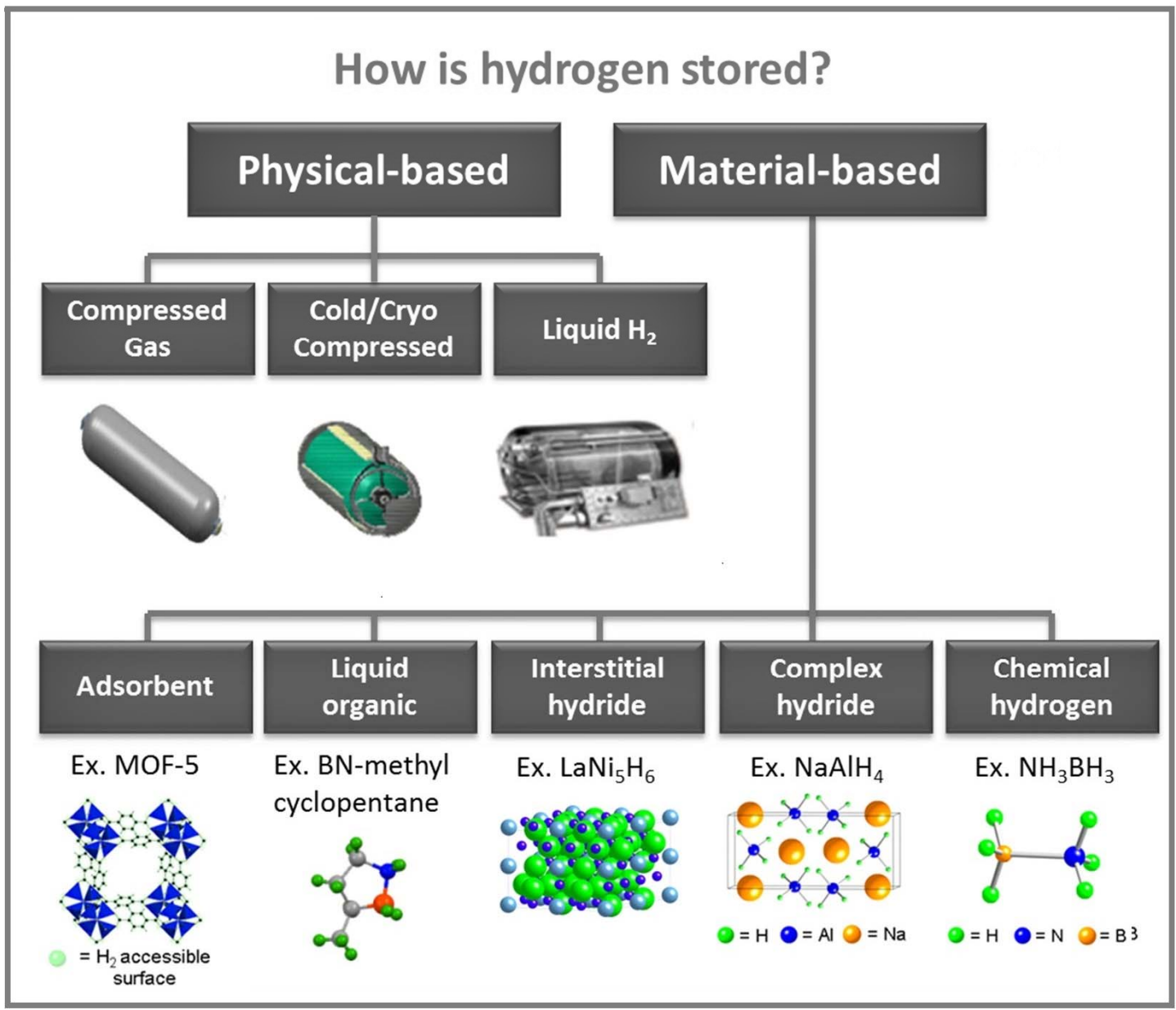

**Figure 3.** Methods to store hydrogen according to the U.S. Department of Energy [14].

However, presently there are major difficulties for transportation, storage, handling, safety, and consumer/user knowledge of hydrogen. The large-scale infrastructure for hydrogen storage and transportation and end-user distribution network is not presently available. The major obstacles for the commercial acceptance and uses of hydrogen as a renewable energy source include:

3.1.(A). Technically demanding leak-proofing of hydrogen pipeline, because of the small size and high mobility of the hydrogen molecule. The development of the no-leak systems substantially increases storage and distribution costs. Issues associated with hydrogen storage and distributions are currently discussed as a barrier to be leaped over for assuring the process implementation [1, 15]. Dedicated hydrogen pipelines may be preferred. Some industrial hydrogen pipelines have existed for decades and could be refurbished to accommodate such hydrogen consumption. However, the effect of the change of gas composition and characters in those pipelines must be fully studied and assessed.

3.1.(B). Alternately, hydrogen can be transported from the production facility to end-users over the road in cryogenic liquid tanker trucks, by compressed gaseous tube trailers, by rail or by barge. Hydrogen used in portable or stationary applications can be delivered by iso-cylinder truck to a storage facility, or in a battery of cylinders. Hydrogen used in fuel-cell electric vehicles (FCEVs) is dispensed very much the way as gasoline is. Nevertheless, the integrity and design of containers and storage cylinders as well as transfer hoses and nozzles must be examined and possibly improved.

3.1.(C). Locations of dispensing stations must be close to end-users while having sufficient buffer and safety measures. In some countries, the hydrogen dispensing stations are located within gasoline stations.

## 3.2. Ammonia ($NH_3$) Production through Renewable Energies:

Ammonia is a compound of nitrogen and hydrogen with chemical formula $NH_3$. Chemical formula and the trigonal pyramidal molecular structure of ammonia are given in Figure 4. The H-N bond angles are 107° apart because the H-atoms are repelled by the lone pair of electrons on the N- atom.

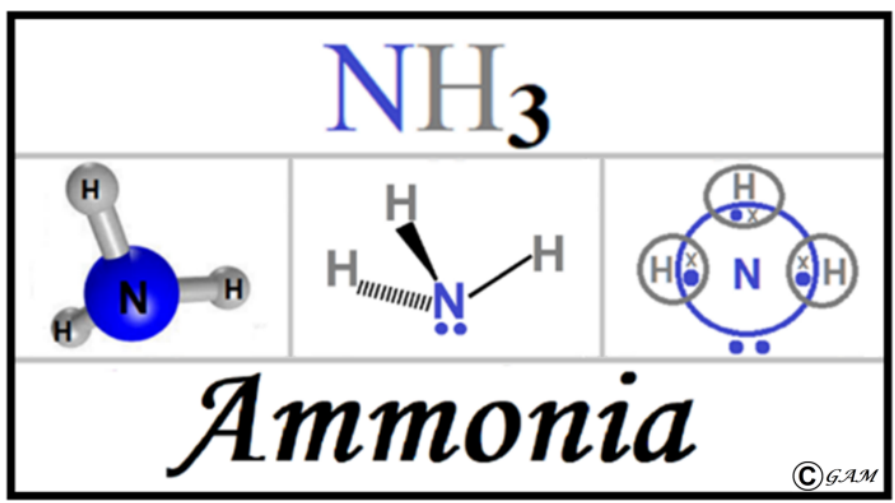

**Figure 4:** Various types of molecular and chemical structure of ammonia molecule as being reported

Ammonia has a molar mass of 17.031 g/mol. At normal conditions it is a colorless gas phase with the density of 0.73 kg/m³. Ammonia boiling point is at -33.34 °C, its melting point is at -77.73 °C, its critical temperature is at 132.22 °C, its critical pressure is at 11.345 MPa, and its critical density is at 234.99 kg /m³ [16].   Ammonia possesses a characteristic unpleasant and toxic odor. It is a polar compound with dipole moment 1.5 Dyne. In aqueous solution, it acts as a weak base and it is used as a general-purpose cleaner and disinfectant.

Several "research and development" groups are proposing to produce ammonia ($NH_3$) using renewable energies, and then store, transport, and consume ammonia [17-20].

Certain organizations worldwide concerned about global warming are investing in the so-called "ammonia energy" technology [20]. The technology for ammonia application as an alternative fuel to replace conventional fuels intended for transportation, commercial and industrial sectors need a great deal of effort to be developed. The technology for ammonia storage and transportation is already available. At present there are many applications for ammonia in all sectors of societies as reported in Table 1.

**Table 1.**  A list of applications of ammonia

**Ammonium nitrate fertilizer**
- Food crops
- Biofuels crops

**In household cleaning products**

**As working fluid in:**
- Absorption cooling cycles (*Refrigeration and Air conditioning*)
- Low T-gradient power cycles,
- Ocean thermal energy conversion.

**Raw material for manufacturing of**:
- Certain plastics,
- Certain fabrics,
- Dyes,
- Explosives,
- Pesticides.

**Applied for treatment of:**
- Water supplies,
- Wastewater,
- Waste.

**Applied in these industries:**
- Beverage & Food,
- Cold storage,
- Medical
- Paper & Pulp,
- Rubber.

**Proposed for future**
- Ammonia as fuel for renewable energies.
- Conversion of ammonia to hydrogen

At present about 90% of the ammonia produced is used for the production of fertilizers. With the expected rapid expansion of biofuels production around the world this percentage of ammonia use as a renewable energy will increase appreciably. At present, the industrial-scale ammonia production is exclusively based on the use of fossil fuels.  Such industries emit more greenhouse gases than any other chemical-making process [21].
As a matter of statistics, about 176 million metric tons of ammonia ($NH_3$) was produced worldwide in 2019 using fossil fuels.  For that, about 500 million metric tons of greenhouse gases were emitted, which amounted to 1.8% of global greenhouse gases emissions [22]. It is estimated that global ammonia production currently contributes almost 11% to industrial carbon dioxide emissions.

Infrastructures for ammonia storage and transportation are already available. For example, USA presently has about 2,000 miles of ammonia transport pipelines, as of December 31, 2018 owned and operated by NUSTAR ENERGY L.P. transporting about two million metric tons of ammonia per year [23]. Another 1,100 miles ammonia pipelines (known as the Magellan Midstream Pipelines) that could have raised the U.S. ammonia pipelines to a total of about 3,100 miles network is currently being decommissioned [22]. The operators of the Magellan Midstream Pipelines have cited concerns about low profit margin, high operating costs and potentially lower ammonia supplies [24].

The use of renewable energies in the production of ammonia is highly desirable in the context of reducing the global emission targets of greenhouse gases (GHGs). The production rate of ammonia for the previous decades (1947-2019) is reported in Figure 5. According to the curve in Figure 5 the growth rate of ammonia production around the world followed an almost exponential trend up to 1986, and then exhibited a stepwise increase.

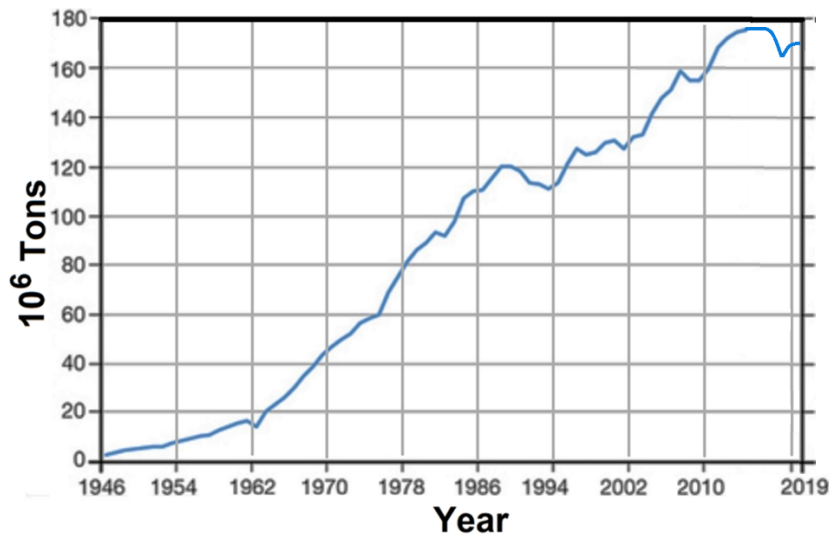

**Figure 5:** Global production of ammonia between 1947 and 2019 in metric tons. Data of up to 2014 was taken from [18].

The following procedure may be used to produce ammonia from renewable energies; its storage and transportation for long distances (see also Figure 6 for details):

3.2.1. A renewable energy is used to dissociate water through electrolysis to produce hydrogen ($H_2$).

3.2.2. Nitrogen ($N_2$) is separated from air (which contains ~80% nitrogen) also using renewable energies.

3.2.3. Hydrogen and nitrogen are reacted to produce ammonia ($NH_3$). This reaction being exothermic produces heat which may be used to produce steam in order to run a power generator for electricity generation to power the ammonia transmission pipeline pumps.

Figure 6 shows the details of production, storage, and transport of ammonia from renewable energy source locations to end-uses.

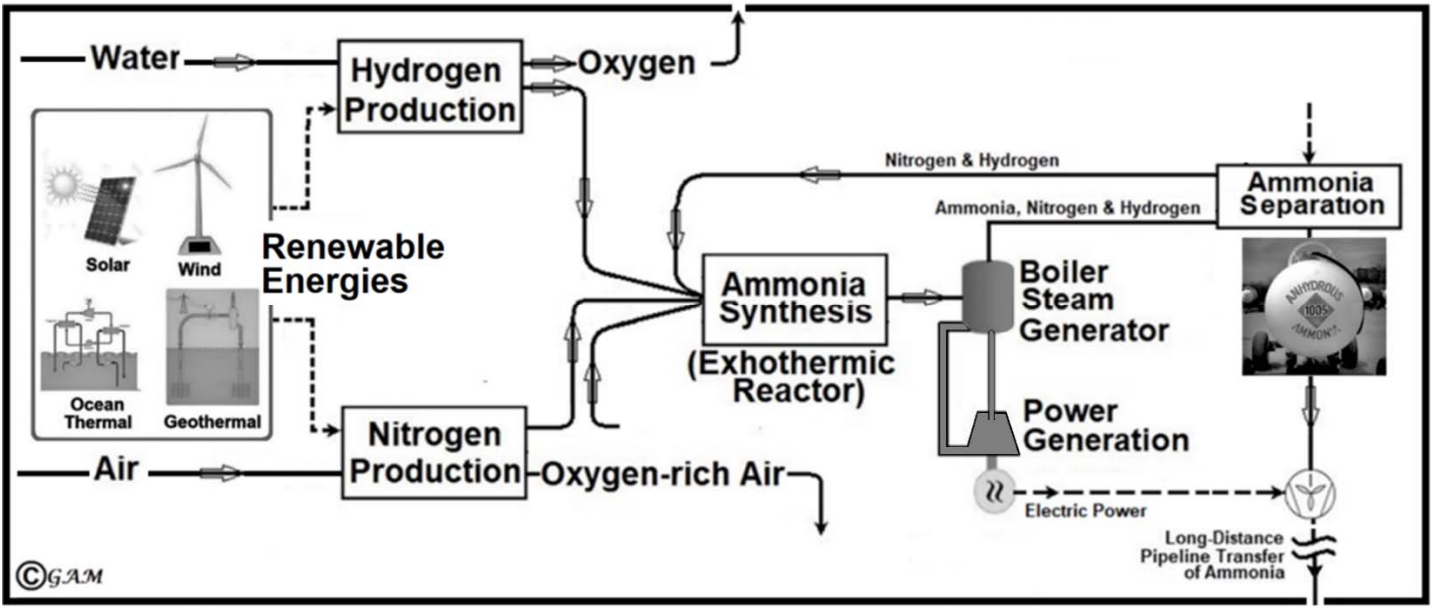

**Figure 6:** General scheme for production, separation, storage and transportation of ammonia utilizing renewable energy

Note that 91.8 kJ heat per mole of nitrogen is released during the exothermic "Ammonia Synthesis" stage [25] of Figure 6:

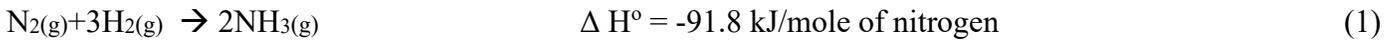

$N_{2(g)} + 3H_{2(g)} \rightarrow 2NH_{3(g)}$ $\quad\quad\quad$ $\Delta H° = -91.8$ kJ/mole of nitrogen $\quad\quad\quad$ (1)

As a result, the released heat can be used to boil water, produce steam, and run a turbine, either to produce electricity or use the resulting mechanical energy directly to pump ammonia for pipeline transfer.

In case ammonia is intended to be used as a fuel to produce heat through combustion in furnaces and in internal combustion engines [26], ideally the following reaction is expected to occur:

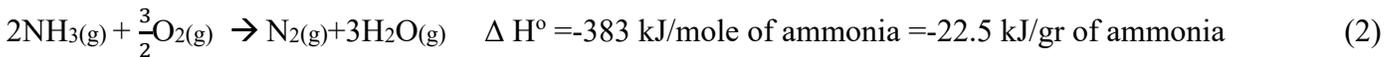

$2NH_{3(g)} + \frac{3}{2}O_{2(g)} \rightarrow N_{2(g)} + 3H_2O_{(g)}$ $\quad$ $\Delta H° = -383$ kJ/mole of ammonia $= -22.5$ kJ/gr of ammonia $\quad$ (2)

However, it is expected that nitrogen oxides (NOx) will also be produced, and the nature and amount of such oxides are a function of reaction conditions [27].

Table 2 shows the standard heats of combustion and normal density (ρ) of ammonia, as compared with other well-known fuels (all exothermic processes).

**Table 2:** Standard heat of combustion of ammonia as compared with other well-known fuels

| Fuel | Combustion Reaction Process | ΔH° | | Density |
|---|---|---|---|---|
| | | [kJ/mol][28] | [kJ/gr][28] | [kg/m³] |

| Ammonia | $2NH_3(g) + \frac{3}{2}O_2(g) \to N_2(g) + 3H_2O(g)$ | -383 | -22.5 | 0.73 [16] |
| --- | --- | --- | --- | --- |
| Graphite | $C(s) + O_2(g) \to CO_2(g)$ | -394 | -32.8 | 2,260[29] |
| Methane | $CH_4(g) + 2O_2(g) \to CO_2(g) + 2H_2O(g)$ | -891 | -55.7 | 0.657[4] |
| Ethanol | $CH_3\text{-}CH_2\text{-}OH(l) + 3O_2(g) \to 2CO_2(g) + 3H_2O(g)$ | -1367 | -29.7 | 789[29] |
| Octane | $C_6H_{18}(l) + \frac{21}{2}O_2(g) \to 6CO_2(g) + 9H_2O(g)$ | -5470 | -47.9 | 703[29] |
| Hydrogen | $H_2(g) + \frac{1}{2}O_2(g) \to H_2O(g)$ | -286 | -143 | 0.0899[30] |

As seen from Table 2, the heat of combustion of ammonia per unit mass is much smaller (in absolute value) compared to the other well-known fuels.

Since hydrogen is a better fuel for fuel-cells, internal combustion engines and many other applications, but its transportation and storage facilities are not widely available, or costly to install, it is mentioned that the ammonia produced through the use of renewable energy sources can be stored and transported to the consumption sites and there it may be converted to hydrogen for energy use. The following reactions can be applied to convert ammonia to hydrogen and combustion of hydrogen to produce water:

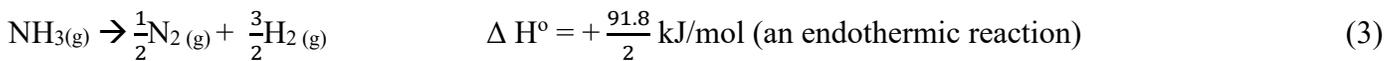

$$NH_3(g) \to \tfrac{1}{2}N_2(g) + \tfrac{3}{2}H_2(g) \qquad \Delta H^\circ = +\tfrac{91.8}{2} \text{ kJ/mol (an endothermic reaction)} \qquad (3)$$

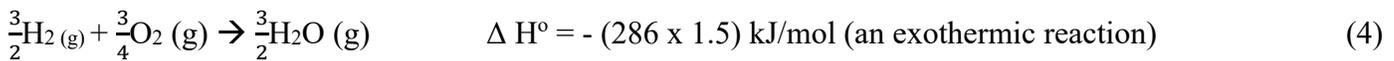

$$\tfrac{3}{2}H_2(g) + \tfrac{3}{4}O_2(g) \to \tfrac{3}{2}H_2O(g) \qquad \Delta H^\circ = -(286 \times 1.5) \text{ kJ/mol (an exothermic reaction)} \qquad (4)$$

The net energy which will be available from the above two reactions after combustion of hydrogen will be:

$$+\tfrac{91.8}{2} \text{ kJ/mol} - (286 \times 1.5) \text{ kJ/mol} = -383.1 \text{ kJ/mol}. \qquad (5)$$

However, the theoretical adiabatic efficiency for the thermocatalytic dissociation of ammonia is about 85% relative to the energy of the released hydrogen. If no other energy source were available, at least 15% of the available hydrogen energy content would have to be burned to supply the heat of reaction [16],

Then, according to Thomas and Parks (2006), $NH_3$ dissociation producing $H_2$, then $H_2$ combustion will have a net energy of

$$(-383) \times (0.85 - 0.15) = -268.1 \text{ kJ/mol} = -15.75 \text{ kJ/g of ammonia}, \qquad (6)$$

which is quite small as compared to other fuels listed in Table 2. In other words, ammonia, as a carbon-free carrier, possesses a rather low energy density. Then ammonia dissociating into hydrogen leads to loosing even more energy. Also considering the efficiencies of renewable energy sources to produce hydrogen through electrolysis and the nitrogen separation from the air, as reported in Figure 6, there will be a great deal of energy loss from source to final hydrogen fuel energy.

Technology for ammonia production using fossil fuels was developed about 100 years ago, to produce fertilizer for increasing food production and support the growing world population in the 20th century. In the 21st century, ammonia production using renewable energy can play a significant role in reducing the greenhouse gases (GHG) emissions. This may contribute to mitigating climate change, due to the increased use of ammonia as fertilizer for biofuels production.

The major application of ammonia around the world, so far, has been to produce fertilizers for food crops. With the advent of biofuels, this important role of ammonia will expand appreciably, if the technology for its production from renewable energies, including its existing infrastructure for storage and transportation is further developed. Bulk storage of ammonia in liquid form can easily be achieved either through refrigeration at -33°C or liquefaction at a pressure of 1-1.5 MPa [31], making ammonia a more versatile compound for renewable energy storage. Also, the existing distribution network around the world allows ammonia to be stored in large, refrigerated tanks and transported by roads, pipelines and ocean tankers.

Current research on potential applications of ammonia as fuel includes utilization in industrial furnaces, gas turbines and fuel-cells for power generation [32]. It has been demonstrated that by co-firing hydrogen or ammonia with natural gas or coal, power generators could systematically reduce carbon emissions from existing generating assets, avoiding significant capital investments [33]. However, application of ammonia for combustion as a fuel source has many problems, including:
- It has a very unpleasant smell even at low emission concentrations.
- It also has toxicity at high concentrations.
- It has a very low energy per unit mass for use as a combustion fuel.
- It has a rather low efficiency of dissociation for hydrogen production.

Production of ammonia from renewable energies, its use as hydrogen carrier and its dissociation to produce hydrogen for fuel are rather energy intensive. Hence, a significant fraction of the original renewable energy collected to produce ammonia is lost in this process. Of course, further research on this subject may produce promising approaches for "ammonia energy" technology.

Application of ammonia as fuel in internal combustion engines presently faces some major challenges [26], including its rather low flammable characteristics, very high fuel NOx emission, and lack of knowledge of the dynamics and chemistry of its combustion. Of course, research on this subject is in progress at various laboratories. However, our present knowledge indicates that the best role for ammonia produced via renewable energy processes will be for the applications shown in Table 1, and its production expansion for use as fertilizer for the cultivation of advanced $2^{nd}$-$4^{th}$ generation biofuels. It is worth mentioning that hydrogen and ammonia also have variety of medical applications, including as geroprotectors [34].

The large-scale infrastructure for ammonia transportation, storage and end-use distribution network is available and has been in use for many years. The following are the safety and environmental aspects of ammonia [35]:

3.2.(A). Swallowing ammonia can cause burns to the mouth, throat, and stomach and can be fatal. Skin contact with ammonia can cause redness, pain, irritation, and burns. It is hazardous when released in large quantities due to its toxicity.

3.2.(B). The hazards and risks associated with the truck transport, storage, and dispensing of anhydrous ammonia are similar to those of gasoline and LPG.

3.2.(C). Ammonia is environmentally benign, having zero GWP (global warming potential) and zero ODP (ozone depletion potential).

## 4. Biofuels Production through Renewable Energies

Biofuels are liquid, solid, or gaseous fuels made from renewable biological materials and are one of the more promising forms of energy for the replacement of non-renewable energy sources (fossil fuels). Global production of biofuels has been growing steadily over the last two decades from 9.2 million metric tons of oil equivalent in the year 2000 to 95.4 million metric tons of oil equivalent in the year 2018 [36]. In the year 2017 biofuels provided roughly 3.8% of total road transport fuel globally [37]. Biofuels development relies heavily on country-specific programs or mandates & outlooks for transportation fuels.

Biofuels can be generally categorized into primary (represented by unchanged biomass) and secondary (represented by processed biomass). Wood, animal fats, and forest residue constitute primary biofuels, while biogas, biohydrogen, and biodiesel represent secondary biofuels. Secondary biofuels are typically divided into four main classes [1,5,38-40].

**First-generation** (conventional),
**Second-generation** (cellulosic)
**Third-generation** (advanced).
**Fourth-generation** (microorganism metabolism)

**The first-generation biofuels** are derived from harvest feedstock, such as sugarcane and corn, which are unsustainable in the long-term and have intense competition with the food industry. These biofuels include ethyl alcohol (ethanol) produced from the fermentation of carbohydrates (from corn or other edible grains) and sugars using enzymes and microorganisms (yeast), biodiesel which is a product of trans-esterification of vegetable oils or fats and green-diesel produced from canola oil using fractional distillation [1,41,42].

**The second-generation (or cellulosic) biofuels** are fuels made from lignocellulosic materials, containing large amount of cellulose. Cellulose, a complex carbohydrate, along with hemicelluloses and lignin are the organic matter found in plant walls. Lignocellulosic feedstock includes agricultural residues (e.g., corn stover), forestry residues (e.g., wood chips), dedicated energy crops (e.g., switchgrass, hybrid poplar), and urban sources of waste (e.g., municipal solid waste), which have minor competition with the food industry. Lignocellulosic biofuels technology is expected to reduce greenhouse gas (GHG) emissions by at least 60%, compared to the 2005 petroleum baseline [43, 44]. A recent scientific data has shown that this estimated threshold by EPA has already been achieved due to technology improvement [44]. Also, it has been estimated that ethanol derived from corn, sugarcane, corn stover, switchgrass and miscanthus can reduce life-cycle GHG emissions by 19–48%, 40–62%, 90–103%, 77–97% and 101–115%, respectively, compared to petroleum gasoline [45].

**The third-generation biofuels** are related to microorganism metabolism, especially algal biomass, and microbes, and could be linked, to a certain extent, to the utilization of $CO_2$ as a feedstock. Microalgae is a feasible and promising source of renewable fuels, since it can produce a wide range of biofuels, biomethane, biohydrogen, and biodiesel. Besides, microalgae harvest is carried out at a high rate and on short harvest cycles, and it does not require large land areas [1,46].

**The fourth-generation biofuels**, like the third-generation one, are based on microalgae processing, focusing on metabolic engineering of photosynthetic microorganisms for producing biofuels [47].

Due to the large application in transportation activities, liquid biofuels are the most desired [1,5]. The use of liquid biofuels as an energy source in the USA dates back to the years when oil lamps and cooking oil stoves and heaters were used as the general means of lighting, cooking, and heating. All types of vegetable oils such as castor, rapeseed, peanut; animal oils, especially whale oil and tallow from beef or pork; refined turpentine from pine trees; and alcohols, especially wood alcohol such as methanol (methyl alcohol) and grain alcohol, such as ethanol (ethyl alcohol) were utilized. The more commonly used fuel in the U.S. before the use of petroleum fuels was a blend of alcohol and turpentine called "camphene" which simply referred to as "burning fluid" [48]. These were early attempts to use biofuels as transportation fuel.

In 1826, Samuel Morey invented a prototype of the first American internal combustion engine powered by a blend of ethyl alcohol and turpentine [49]. But Morey's work was curtailed by lack of funding and the world's interest directed towards steam engine locomotive using coal [48].

In 1896, Henry Ford built the first automobile (a quadricycle) to run on pure ethanol. But a $(US)2.08 per gallon liquor tax introduced earlier to pay for the American Civil War, which was still in effect at the time of Ford's automobile, impeded the success of his invention. The liquor tax made ethanol more expensive, forcing people to use coal oil (which was known as kerosene) that was taxed only 10 cents per gallon [48,50]. Coal oil and kerosene were often assumed to be the same thing; a clear, liquid fuel used for lamps and cooking. In the early years of the oil industry, the two names were often used synonymously.

After the repeal of the liquor tax in 1906, Ford declared ethanol as the fuel of the future, and in 1908 introduced the Model T vehicle to run on ethanol with gasoline as an option [49,51]. But the euphoria surrounding ethanol as the fuel of the future was only ephemeral. This was because of the discovery of new underground petroleum fields by independent petroleum companies introduced cheap gasoline on the market [48].

Nevertheless, in 1917, Alexander Graham Bell, while addressing the graduating class of the McKinley Manual Training School in Washington, D.C., on the theme "Prizes for the inventor: some of the problems awaiting solution", mentioned alcohol as one source of fuel supply which may solve our future energy problem. According to Bell [52] "alcohol makes a beautiful, clean, and efficient fuel, and were it not intended for consumption by human beings, it could be manufactured very cheaply in an indigestible or even poisonous form". Based on this statement, Bell predicted that the world would probably depend on alcohol more and more as time goes on, with opportunities opening up for engineers who will modify our machinery to enable alcohol to be used as the source of power. Among the possible feedstock mentioned by Bell for alcohol production included sawdust, corn stalks, waste products from farms, garbage from our cities and almost any vegetable matter capable of fermentation [52].

In general, the use of first-generation biofuels based on food crops, such as corn grains, which will have to be produced in excess of what is needed for food supply, can result in an increased rate of deforestation. It should also be mentioned that as farmers are driven by market and profits forces to balance the supply and demand of food crops used for human consumption, they are inadvertently diverting their land for ethanol production, a practice that may become one of the agents of deforestation. This situation can negatively impact the control of greenhouse gas emissions, as forest plants depend on $CO_2$ for photosynthesis. Also, the use of corn-derived ethanol as a biofuel would add about $(US)0.75 to $(US)1.00 per bushel to the price of corn which is the mainstay for millions of people around the world [53]. Given that corn and other agricultural food products need to be grown, collected, dried, and fermented to produce alcohol, and each of these stages requires the use of energy in excess of energy content of the product, the first-generation of biofuels becomes unsustainable and not economical.

The ethanol fuel energy balance (also called "net energy gain"), is the difference between "the energy released by burning the resulting ethanol fuel" and "the total amount of energy input into the process of growing corn and producing alcohol from corn". This difference is marginal [54] meaning that alcohol produced from corn cannot be used economically as a transportation fuel. Additionally, the problem of deforestation is probably inevitable, unless there is an alternative approach for ethanol production as a second-generation (cellulosic ethanol) biofuel, or an advanced (third-generation) biofuel.

The cellulosic (second-generation) biofuels are the derivatives of sustainable feedstock. These include cellulosic ethanol, biohydrogen, biomethanol, BioDMF (DMF=2, 5-dimethylfuran), BioDME (DME=dimethylether) and Fischer-Tropsch fuels. Requirements for the sustainability of a feedstock, as defined, include among others, the continuous availability of the feedstock, its zero or minimum impact on greenhouse gas emissions, its biodiversity, and optimum land use.

Advanced (second-, third- and fourth-generation) biofuels can improve the energy balance efficiency, $\eta_{Energy\ Balance}$, which is defined as,

$$\eta_{Energy\ Balance} \equiv \frac{E_{Biofuel}}{E_{Input}} - 1 \qquad (7)$$

In this equation $E_{Biofuel}$ is the total energy produced by the biofuel, divided by $E_{Input}$, the input energy (energy used in farming, processing, production) needed to produce it. Positive energy balance efficiency means more usable energy is produced than is required to produce the biofuel, while negative energy balance efficiency means the biofuel takes more input energy to produce than it yields.

Several data on net energy balance ratios of fuels are reported in the literature. The most recent studies have concluded that the net energy balance for corn ethanol production averages from 2.1–2.3, while that for sugar cane ethanol ranges from 8.32-10.22 [55-57]. Also depending on the feedstock (switchgrass, miscanthus, corn stover) the net energy balance ratio for cellulosic ethanol fuels ranges from 1.2 – 5.6 [58,59].

In general, corn production requires a significant amount of energy and financial investment. But the last decade saw the declining of energy use for corn production was declining during the last decade while the farm price for corn was increasing. It was estimated that about $1.59 \times 10^6$ kJ per metric ton of energy was required for corn production in U.S. in 2010, representing a decline in 1996 estimate of about $2.76 \times 10^6$ kJ per metric ton [56]. This decrease in energy usage was attributed to improved production methods, such as locating ethanol plants closer to corn farms, saving transportation fuel and time. Contrarily, from 2006-2013, the average farm price for corn was $(US)185.19 per metric ton, compared to farm prices from 2000 and 2005, which averaged $(US)82.80 per metric ton [60]. This rising farm price was due to increased use of corn grains for ethanol production, an undesirable situation that could cause soaring prices of corn-based food products.

While in South America, especially in Brazil, ethanol has been produced from sugarcane, presently most of the ethanol used as transportation fuel in the USA is derived from corn ethanol. The technology for corn derived ethanol production is rather old, and it goes back to the 1970s when U.S. policymakers were looking for a way to mitigate the high cost of petroleum-based fuels due to the energy crisis at that time [60]. The technology for ethanol from sugar cane is also well established and some countries in Central and South America are heavily involved in its production.

The potential for the production of biofuels to have very high energy ratios efficiency worldwide is with the **second-generation** (cellulosic), the **third-generation** (advanced), and the **fourth-generation** (microorganism metabolism) class of biofuel technologies. That could contribute greatly to the future advances of renewable energy need when we end using fossil fuels.

## 5. Available and Potential Biomass Resources

This category of biomass resources includes low-impact crops for biofuel production, algae as potential biomass feedstock, use of abandoned agriculture lands to grow energy crops and urban waste as feedstock for biofuels production.

In this section includes discussion about biomass resource categories, including low-impact crops for biofuel production, algae as potential biomass feedstock, use of abandoned agriculture lands to grow energy crops and urban waste as feedstock for biofuels production.

**5.1. Low-impact crops for biofuels production:**
According to the U.S. Department of Energy (US-DOE), "low-impact" crops are agronomic crops that can sustainably be produced over a broad range of geographic area in the U.S. at a cost that is competitive with other lipid producing crops.

Lipid is referred to as a broad group of naturally occurring molecules including fats, fat-soluble vitamins, monoglycerides and diglycerides, phospholipids, sterols, waxes, and others. The requirement of a minimum amount of water, fertilizer, pesticides, energy, and land area make the low-impact crops much more economical for biofuels production than traditional crops, such as corn and soybean. With no impact on the food sector, low impact crops would not be used for human consumption. Besides, the production of such crops would not cause excessive soil erosion, loss of wildlife habitat, decrease in productivity in other rotational crops produced on the same land, or cause other environmental pollution or health hazards while requiring few inputs thus providing economic, environmental and social benefits" [1,61].

In essence, low-impact crops could be valuable resources for producing high quality biodiesel fuels. The use of low-impact crops can mitigate the competition for food and land, while having the potential to achieve higher quantities of biofuel per acre of land than currently being achieved using corn or soybeans. Examples of "low-impact crops" are agricultural residue such as corn stover (Figure 7), wood chips, straws of cereals such as oats, sorghum and wheat, including high yielding herbaceous energy crops, such as perennial grasses for cellulosic biofuels production. The large availability of biomass residues provided from agriculture and forestry activity point out to a positive economic viability of low impact crops. The largest amount of agriculture and forestry waste is found in the Asia and South America regions, especially related to the high activity of rice and sugarcane harvest [1,5,62,63].

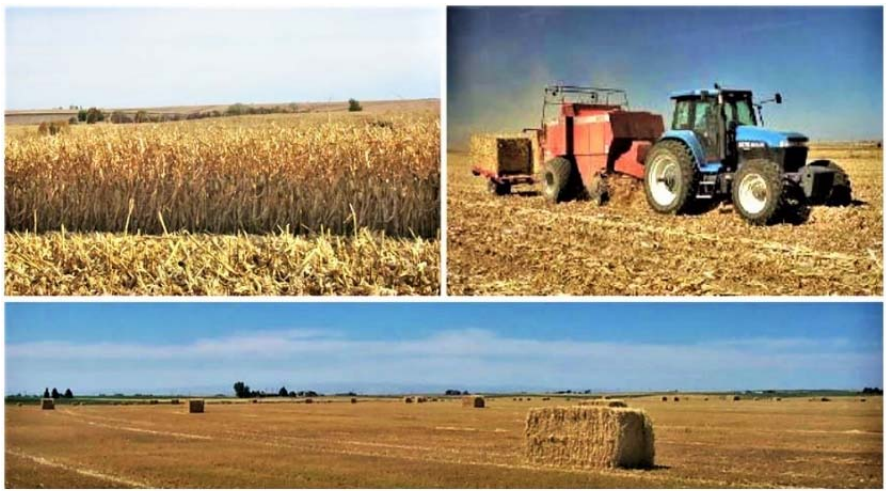

**Figure 7.** Corn stover on the field [64].

Many of the technologies for conversion of agricultural cellulosic waste, like corn stover to ethanol, are well developed. The amount of corn stover from these croplands would be enough to feed ethanol plant with an estimated capacity of 291,000 tons per day, when ethanol price is set at $(US)56.7 per barrel [65]. Hence, the construction of a plant to process 291,000 tons of corn stover per day to produce ethanol requires $(US)2.75 billion investment. The overall economic state impacts will include increased economic activity, the creation of about 20,000 jobs, and moving the state closer to energy-independence [65]. By 2022, the U.S. EPA has projected that corn stover in Illinois can contribute about 34.38 million barrels of cellulosic ethanol per year to meet the EISA standard [66].

**5.1.1. *Woody biomass*:** These types of biomass are categorized as forest residues, primary mill residues, secondary mill residues and urban wood residues (see Figure 8). Since woody materials have high energy densities, they can produce large amount of biofuel per ton of feedstock. The biomass energy content can be assessed by the heat value, categorized as higher heating value (HHV) and lower heating value (LHV). Cellulose-rich materials are more prone to be processed using thermal degradation methods; however, many woody biomass residues contain high amounts of lignin. Lignin-rich materials are preferentially recycled by alternative methods or submitted to lignin extraction, which has found valuable applications. In an integrated biorefinery structure, the lignin-rich residues from the biochemical process can also be thermochemically converted into bio-oil, coatings, and chemical intermediates [64,65]. The U.S. EPA has identified several feedstock pathways (i.e., various feedstock supply chain designs for thermochemical biofuel conversion systems) and imports from countries such as Brazil to meet the U.S. EPA's amended Renewable Fuels Standards (RFS2). These feedstocks are defined within the context of the U.S. Energy Independence and Security Act (EISA) of 2007 and include switchgrass, soybean oil, corn oil, crop residues and woody biomass [1,43].

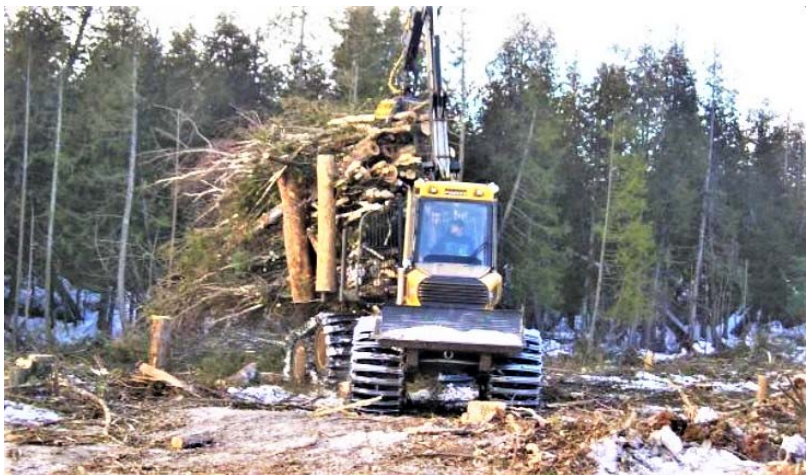
**Figure 8.** Woody biomass [67]

5.1.1.A. *Forest residues*:
This category of biomass includes logging residues and other "removals" from forest resources such as unused portions of trees cut, or trees killed by logging and left in the woods. The rest of the removals are considered trees cut or otherwise killed by forestry operations (e.g. pre-commercial thinning, weeding, etc.) or land clearings and forest uses that are indirectly linked with round wood product harvests. However, the volume of wood removed from the inventory by the reclassification of timberland to productive reserved forestland is excluded.

5.1.1.B. *Primary mill residues*:
These types of biomass consist of crushed wood materials (coarse and fine), as well as barks generated at manufacturing plants (primary wood-using mills), when round wood products are processed into primary wood products, such as slabs, veneer clippings, trimmings, edgings, sawdust and pulp screenings. Recycled byproducts from mill residues, as well as those left unused and disposed of as waste are also included in this category.

*5.1.1.C. Secondary mill residues:*
The secondary mill residues include wood scraps and sawdust from woodworking shops, wood container and pallet mills, furniture factories and wholesale lumberyards. It also includes wood scraps from wood kitchen cabinet and countertop, non-upholstered wood, household furniture, wood office furniture, wood window and door manufacturers and custom architectural woodwork and millwork.

*5.1.1.D. Urban wood residues:*
The three main categories of urban wood residues are: (i) municipal solid waste (MSW), (ii) wood which includes wood chips, pallets, and yard waste, utility tree trimming and/or private tree companies, and (iii) construction/demolition wood. All these woody biomasses typically have high lignin contents. This makes them less attractive feedstock for biochemical conversion, but they are suitable for thermochemical conversion.

**5.1.2. *Dedicated energy crops:*** Renewable energy crops are lignocellulosic materials, specifically developed and cultivated for biofuel production because of their high energy yields. The energy may be generated through direct crop combustion or gasification to produce electricity, or through the crop conversion into liquid fuels such as ethanol for use in the transportation industry. Several promising plant species are identified as suitable for liquid fuels which include miscanthus, switchgrass, hybrid poplars, silver maple, reed canarygrass, black locust and sorghum [68]. In the continental United States, there are several growing zones with variable factors, such as mean temperature, soil quality and rainfall. However, since there is not a single plant species that is optimal for the entire U.S. growing zones, the use of different species as energy crops is a viable approach to obtain sufficient biomass for large-scale production of liquid biofuels [68]. In general, energy crops are classified into two types: (A) Herbaceous energy crops [69]; (B) Short-rotation woody energy crops [70]. In the following sections, we briefly describe the characteristics of each of these energy crops.

*5.1.2.A. Herbaceous energy crops:*

Herbaceous energy crops are mostly the grass species that are harvested like hay. Since these grasses regrow from their roots, they do not require replanting for long periods (15 years or more). They include perennial grasses, such as switchgrass, miscanthus, bluestem, elephant grass and wheatgrass [70]. The current research interest is more focused on switchgrass and miscanthus.

Switchgrass (*Panicum virgatum L.*) is considered the most productive warm-season grass native from the USA to be used as a feedstock for fuel production. Switchgrass is a large prairie grass (Figure 9) which grows in the US Midwest and can be transformed into cellulosic biofuels. Switchgrass normally attains heights of 1-1.5 meters; however, it can grow beyond 3 meters in some southern states in the U.S. Its plant is resistant to many pests and diseases and it has a deep root system extending a few meters below the ground surface, allowing the plant to tolerate many soil and climate types [66].

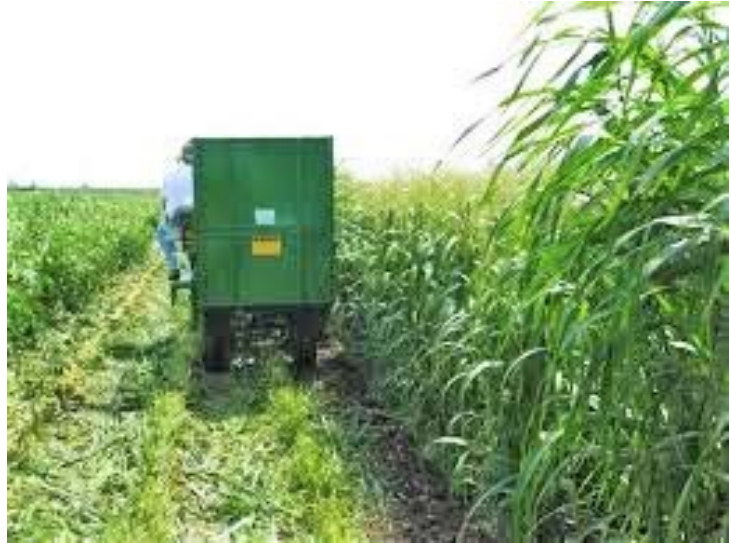
**Figure 9.** Switchgrass [71]

The potential impact of using grass species biomass, like switchgrass, instead of edible biomass such as corn, to produce biofuels has been investigated [1,69,72,73]. The studies point out to a growing of grass species in the US Midwest and in the same cropland area allocated for corn cultivation could increase ethanol production, reduce the nitrogen leaching, and decrease greenhouse gas emission. Also, it has been estimated that the potential switchgrass production on Conservation Reserve Program (CRP) Lands is more than five million dry metric tons per year [74]. CRP is a voluntary program administered by the USDA Farm Service Agency for agricultural landowners. The program provides financial and technical assistance to eligible ranchers and farmers for the management of water, soil and other related natural resources on their lands [74].

Miscanthus giganteus (MG) is a Japanese native hybrid and tall perennial grass which is currently used in the European Union as a commercial energy crop, mainly as a source of heat and electricity, or for conversion into biofuel products such as ethanol (see Figure 10).

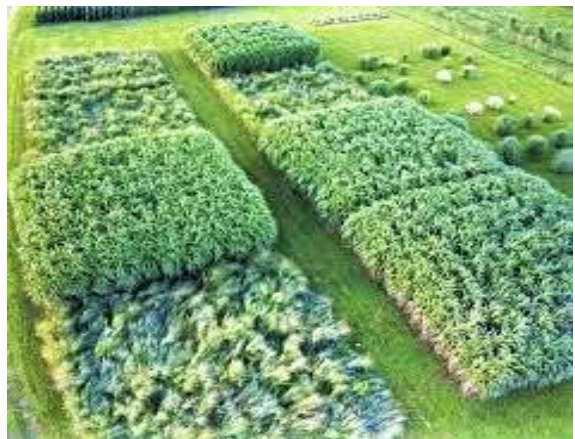
**Figure 10.** Miscanthus giganteus [75].

MG was introduced to Europe in the 1930s as ornamental plant, but its potential yield for cellulose fiber production was not investigated until late 1960s [76]. Subsequent trials for bioenergy production of the plant began in Denmark in 1983, as it spread to Germany in 1987, prior to comprehensive European evaluation. Researchers are studying to determine whether biofuel grasses like miscanthus giganteus could be a viable crop in the U.S. Additionally, the researchers will explore how this viability varies by location [77].

A comparison of the properties of dried herbaceous energy crops (switchgrass and miscanthus giganteus) is shown in Table 3.

Table 3. Properties of dried herbaceous energy crops

| Property | Miscanthus (*giganteus*) | Switchgrass |
|---|---|---|
| Cellulose [%] | 40 - 60 [78,84] | 44-51 [79,80] |
| Hemicellulose [%] | 20-40 [78,84] | 42-50 [79,80] |
| Lignin [%] | 10-30 [78,84] | 13-20 [79,80] |
| Protein [%] | 2.5-5 [80] | < 3.3 [81,85] |
| | | |
| **Harvestable Biomass (Tons/hectare/year)** | 29.6-38.2 [83] | 10.4-12.5 [83] |
| Ash [%] | 1.5-4.5 [82] | 2.8-7.5 [82] |
| Lower heating value in [MJ/kg DM] | 17.8-18.1 [82] | 16.8-18.6 [82] |

[MJ/kg DM] = Megajoule per kilogram dry mass

*5.1.2. B. Short-rotation woody energy crops:*
These are tree crops grown on short rotations, usually with more intensive management than timber plantations. Short rotation forestry refers to the practice of cultivating fast-growing trees that reach their economically optimum size between the ages of 8 and 20 years [86]. They include silver maple, cottonwood, black locust, poplar, and willow [70].

The black locust tree (*Robina pseudoacacia*) is a nitrogen-fixing legume plant that is native to Southern Europe and some parts of the United States [87]. It is considered a promising energy crop for biofuels production. There are more than six ways in which black locust can be utilized as a fiber crop or in producing large amounts of biomass at relatively low energy inputs. These include leaves and young stems for fodder, pulp for paper, leaves and young stems for solid, liquid, or gaseous fuels and extraction of specialty chemicals, such as natural wood additives [88]. The initial studies on black locust have shown promising results because the species produced more biomass material than the next closest species by almost three times. Additionally, the black locust has a rapid growth rate relative to other woody biomasses. Based on these initial results, it is estimated that black locust production may have the potential as an energy crop for the US Midwest, especially in northern and southern Illinois where the land is unsuitable for corn and soybean production [89].

The high yield rates of hybrid poplars and willows woody crops coupled with their easy adaptations to many growing conditions give them a suitable characteristic to be used as potential energy crops. It has been estimated that the potential willow or hybrid poplar production on Illinois CRP lands is ~4.5 million dry metric tons per year [74]. Hybrid poplars (*Populus* spp.) are fast-growing trees typically grown under intensive control (known as silviculture) and they are used as saw and veneer logs, fiber production for the pulp and paper industry, phytoremediation, and biofuels feedstock production [90]. The hybrid poplars can reach a height of over 20 meters and a diameter of approximately 30 centimeters within 6 years. Some hybrid poplars yield up to 10 tons of dry biomass per acre per year, which is 5-10 times larger than the yields from natural forests. Because of these attributes, scientists continue to crossbreed hybrids to create trees that will grow faster and will be more drought-tolerant and insect-resistant [70]. Willows (*Salix* spp.) are fast-growing shrubs that can reach a height of 1-1.5 meters in the plantation first year, thus being able to produce multiple stems from the stump cuts. After the first year of cultivation, the plant needs to grow for another 3-4 years to about 5-6 meters before it can be harvested. The harvesting costs are normally lower than other woody crops [70]. Also, willow can be planted in temperate climate areas. Other characteristics that make shrub willows ideal feedstock for biofuels, bioproducts and bioenergy are the spread feasibility from dormant hardwood cuttings, ease of breeding for several characteristics, a broad underutilized genetic base, good chemical composition and energy content, as well as the ability to re-sprout after multiple harvests [91]. Table 4 shows a comparison of the properties of short-rotation woody energy crops (hybrid poplar, willow, and black locust).

**Table 4.** Properties of dried short-rotation woody energy crops

| Property | Hybrid poplar | Willow | Black locust |
|---|---|---|---|
| Cellulose [%] | 4–55[92,98] | 54.6[99] | 52.2[99] |
| Hemicellulose [%] | 24-40[92,98] | 17-19[95] | 18.1[99] |
| Lignin [%] | 18-25[92,98] | 13.3[99] | 16.9[99] |
| Protein [%] | 20-23[96] | 10-13[96] | 22-24[96] |
| Yield [$t_{dm}$/ha/year] | 10-20[93] | 6-10[93] | 8.1-9.7[100] |
| Ash [% dry weight] | 1.43-2.1[94,99] | 2-2.9[93,99] | 1.7[99] |
| Lower Heating value in [MJ/kg DM] | 18.564[97] | 18.599[97] | 21.19[97] |

[MJ/kg DM] = Megajoule per kilogram dry mass
[$t_{dm}$/ha/year] = Dry mass yield in metric tons per hectare per year

5.1.2.C. *Pennycress*:

Pennycress is an oily seed crop used as feedstock for industrial uses. It is a member of the mustard plant family that can produce seeds containing (20-36)% of oil, with 38.1% erucic acid ($CH_3(CH_2)_7$-CH=CH-$(CH_2)_{11}$-COOH) as the major fatty acid [101]. In Figure 11, field pennycress and perfoliate pennycress are shown.

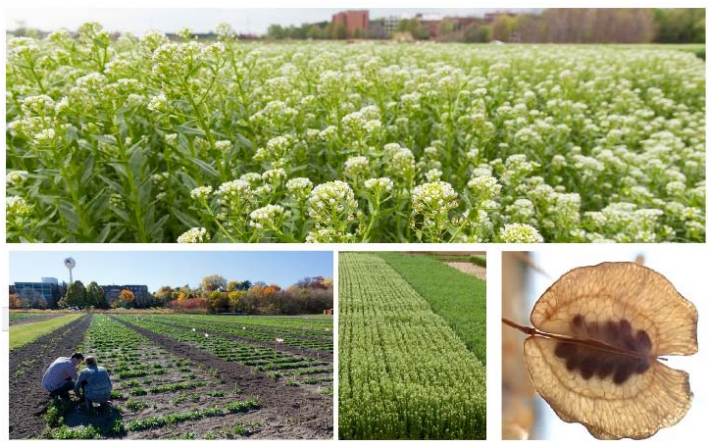

**Figure 11.** Field pennycress [102]

The chemical composition and properties of pennycress oil are presented in Table 5. Data in Table 5 make pennycress ideally suitable for conversion into biodiesel or green renewable jet fuels. The cloud point of pennycress biodiesel (characterized as a methyl ester mixture) is -10°C and the pour point is -15°C, similar to those of soy-based biodiesel which are -9°C (for cloud point) and -12 to -16°C (for pour point) [103,104]. Pennycress biodiesel is less saturate than soy-based biodiesel, making it more resistant to oxidation [101]. Also, on the same acre of land, pennycress seeds are known to yield two times more oil than soybeans.

**Table 5.** Composition and physical properties of pennycress and its oil

| Chemical composition | Seed | Press cake* |
|---|---|---|
| Moisture [%] | 8.8[105] | 8.4[105] |
| Oil [wt%] | 29-36[105,106] | 10.7[107] |
| Crude Protein [%] | 19.6[105] | 25.8[105] |
| | | |
| **Physical Properties** [104] | Oil | Methyl ethers** |
| Pour point [°C] | -18 | -15 |
| Cloud point [°C] | -10 | -10 |
| Flash point [°C] | 234 | 136 |
| Viscosity Index | 222 | 277 |

(*). *A press cake or oil cake is the solid left behind after pennycress seed is pressed to extract the liquids*
(**). *Methyl ether is a light volatile ether (CH3OCH3) obtained by etherification of methyl alcohol*

Pennycress plant can produce between 1.5 - 2.0 tons of seed per acre of land to yield approximately 1.79 - 2.38 barrels of biodiesel [108], resulting in a potential to produce from 75 to 115 gallons of biodiesel per acre [101, 108] (equaling to 701 to 1075 liter per hectare). According to the U.S. Department of Agriculture (USDA), additional biofuels can be obtained from the pennycress press cake using pyrolysis, which increases the yield to about 3.57 barrels per ton of seeds [101]. Therefore, USDA has indicated that pennycress has the potential to produce nearly 142.9 million barrels per year of U.S. liquid transportation biofuels, which will require approximately 40 million acres of land annually [61].

In USA, pennycress is grown as a non-food crop in the winter, when Midwest corn-belt land is normally left uncultivated, that is between the fall corn harvests and before spring soybean and corn planting seasons [101]. Once planted in the winter, the pennycress seed remains dormant till late in the fall before germinating, because of the insufficient light able to stimulate germination during the winter [109]. In the winter, the leaves of the plant start growing close to the ground, which prevents soil erosion. By late April and May in spring, the plant begins to flower, reaching a height of about 0.75 meter and is ready for harvesting in early June [101].

5.1.2.D. Industrial Hemp

Industrial hemp (hemp) is a plant that originated in Central Asia, belonging to the Cannabis *Sativa* species (Figure 12). Hemp is often confused with *cannabis* (marijuana), which also belongs to the same species. The U.S. 2018 Farm Bill showed that hemp can be distinguished from marijuana. On one side, hemp must contain no more than 0.3% concentration of delta-9 tetrahydrocannabinol (delta-9 THC), the primary psychoactive chemical of marijuana. On the other hand, about 1% THC concentration is generally considered as the threshold for cannabis (marijuana) in the US [110]. Industrial hemp is purposefully grown for its fiber and seeds. Applications for the industrial hemp bast fibers include specialty textiles, paper and composites used as an alternative to fiberglass in the automotive and aviation industries. Also, the press cake of hempseed can be added to various foods and beverages, including salad and cooking oil [111]. It must be recalled that in 1941, Henry Ford, the automaker, introduced a prototype car made of hemp composites, and powered by 100% of hemp ethanol [112]. But despite the benefits of industrial hemp, cultivation of the plant was not allowed in the US from 1937-2018, requiring permission from the US Drug Enforcement Agency (USDEA) [113,114]. Currently, the growing interest in carbon-free energy sources has made industrial hemp one of the potential biofuel energy sources under consideration. This is because both cellulosic ethanol and biodiesel can be produced from hemp fiber and hempseed oil, respectively. It is estimated that hempseeds can produce almost four times more oil per acre than soybeans [115]. Table 6 shows the chemical composition, physical properties and the biofuels derived from industrial hemp.

Hemp usually grows in temperate climates, but it can also be cultivated under most climatic conditions in almost every soil with minimum water and fertilizer content. It is an annual plant cultivated from seeds that can grow 5 meters high, requiring average monthly rainfall not less than 65 mm throughout the entire growing season [116]. Hemp plants that are grown purposefully for fiber are densely cultivated and harvested at an average height of 3.0 - 4.5 meters, yielding about 1-5.5 metric tons per acre of dry matter (whole dry stems) [110]. On the other hand, hemp plants cultivated solely for seeds are harvested at an average height 2 - 3 meters, with an average yield of ~350-450 kg per acre, reaching up to 725 kg per acre. The 2017 farm price for hemp fiber was $(US)70-135 per metric ton, while seed price was $(US)1.45-1.65 per kg [110].

**Table 6.** Chemical composition and physical properties of hemp and hemp-derived oil.

| Chemical Composition | Seed [117] | Press cake [118] |
|---|---|---|
| Moisture [%] | 8.1 | 6.35 |
| Oil [wt %] | 25-35 | 9.53 |
| Crude protein [%] | 20-25 | 32.31 |
| Carbohydrate [%] | 20-30 | 2.8 |
| Crude fibers [%] | 10-15 | 43.87 |
| | | |
| **Physical Properties [119]** | Brown hemp Oil (*) | Brown hemp methyl ester (+) | Brown hemp ethyl ester (#) |
| Pour point [°C] | -10 | -17 | 3 |
| Cloud point [°C] | 2 | -4 | 8 |
| Flash point [°C] | 125 | 47 | 60 |

(*) *Brown hemp oil – Hemp seed oil used as biodiesel with no modification*
(+) *Brown hemp methyl ester – Hempseed oil converted to biodiesel through transesterification with methanol*
(#) *Brown hemp ethyl ester - Hempseed oil converted to biodiesel through transesterification with ethanol*

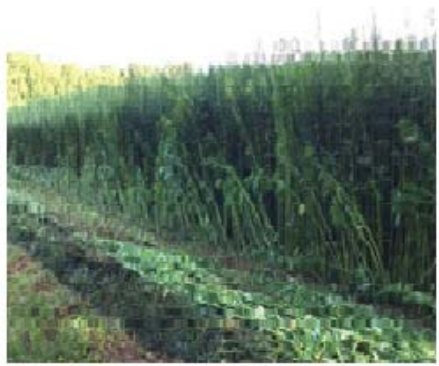
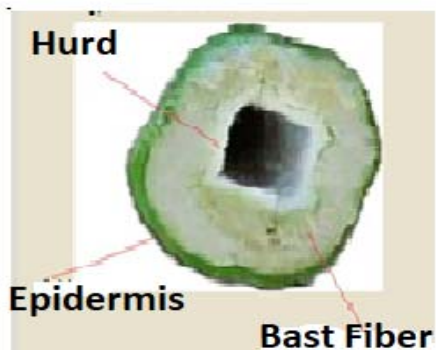
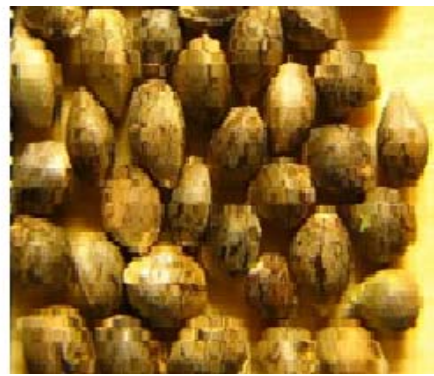

**Figure 12.** Left: Hemp field being harvested for fiber production. [Kentucky Department of Agriculture's Industrial Hemp Program photo) [www.farmanddairy.com/uncategorized/whats-next-for-industrial-hemp-growers/561145.html]; Middle: Anatomy of Hemp; Right: Hempseeds [iowahia.org/anatomy-of-hemp/]

## 5.2. Algae as a potential biomass feedstock for biofuels production:

The technology for macro- and micro-algae cultivation as biomass and its contribution to greenhouse gas reduction in the atmosphere is well established [1,120-122]. Algae can efficiently convert carbon dioxide into biomass under solar radiation. Probably algae can achieve higher biomass production rates, compared to land-based crops based on the land surface area. Macro- and micro-algae can grow in any kind of aqueous system, while utilizing photosynthesis to generate biomass. Macro- and micro-algae can either be heterotrophic or autotrophic. Autotrophic algae need inorganic compounds such as carbon dioxide and a light energy source for growth [122]. Heterotrophic algae, however, are non-photosynthetic and require an external nutrient for growth. But some photosynthetic algae are mixotrophic, in that they can acquire exogenous organic nutrients and perform photosynthesis as well [122].

Macro-algae, also known as seaweeds, are fast-growing multicellular plants that can grow to a size of up to 60 meters in length. Based on their pigmentation, these plants are classified into three broad groups: brown seaweed (*Phaeophyceae*), red seaweed (*Rhodophyceae*) and green seaweed (*Chlorophyceae*). Seaweeds are mainly used for food production and the hydrocolloid extraction [123]. Micro-algae are microscopic thallophytic plants (i.e. plants without leaves, stems and roots) [121]. The three most important classes of micro-algae relative to their abundance are the diatoms (*Bacillariophyceae*), the green algae (*Chlorophyceae*), and the golden algae (*Chrysophyceae*), such as shown in Figure 13.

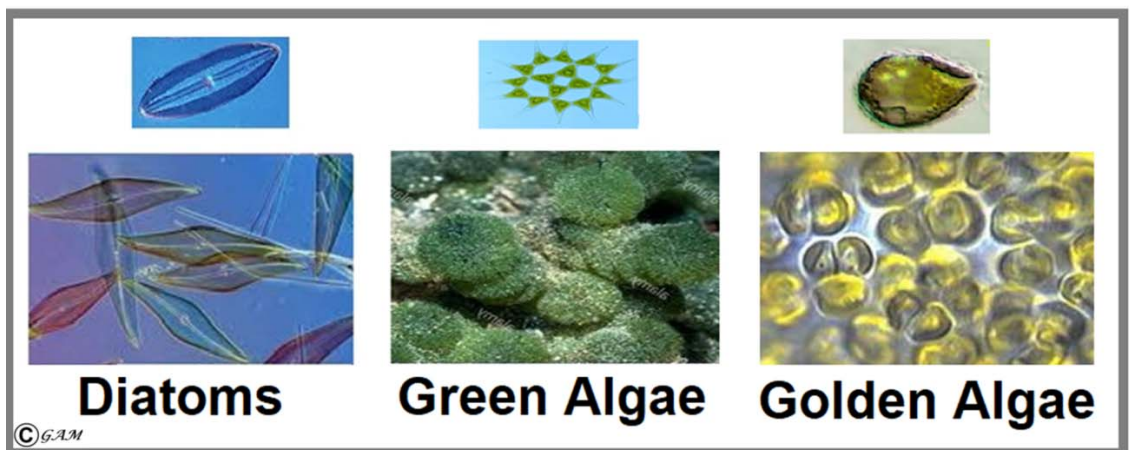

**Figure 13.** The three most important classes of micro-algae

Diatoms, which are the dominant species of phytoplankton, may be the largest group of biomass producers on the earth. Green algae are especially abundant in fresh waters. Green algae store energy mainly in the forms of starch and lipids. The golden algae are similar to the diatoms and they produce oils and carbohydrates [123]. Microalgae are capable of producing hydrogen, ethanol, and biofuels (Figure 14). Their metabolic pathways are coordinated through complex mechanisms that regulate photosynthetic output for the synthesis of proteins, nucleic acids, carbohydrates, lipids, and hydrogen [1,124,125]. Furthermore, micro-algae have broad bioenergy potential, since they can be used to produce liquid transportation fuels, including biodiesel, bioethanol, heating fuels and biojet fuels. Based on land surface area use, micro-algae are capable of producing 15-300 times more oil for biodiesel production than traditional crops [121].

There are several species of algae capable of accumulating large amounts of oils that can be converted into biodiesel through a trans-esterification process. A collection of about 3,000 strains of algae that have been screened for their oil-producing potential is available [126]. *Nannochloropsis sp.* and *Nitzschia sp.* are promising examples of algae, having oil contents of 68% and 47%, respectively [123,127]. However, energy storage of their feedstock can be hampered by the presence of polyunsaturated fatty acids (PUFAs) and high moisture contents which cause oxidation reactions. Besides, the challenges of cost-effective large-scale production of algae biomass have slowed down the pace of their commercial-scale conversion into energy.

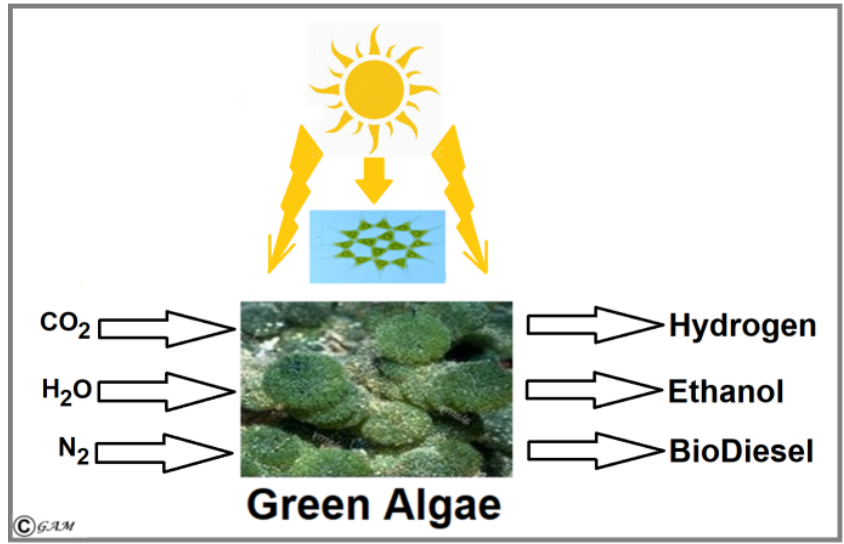

**Figure 14.** A symbolic example of the production of hydrogen, ethanol, and biodiesel by green algae under solar radiation. Details of the mechanisms that regulate such photosynthetic outputs are well described elsewhere [124,125]

Table 7 shows the comparison of the properties of algal-derived bio-oil (green crude), wood-derived bio-oil and petroleum. Similar to other biomass, the main advantages of algal bio-oil over petroleum are that it is renewable, biodegradable, and non-toxic. In addition, the combustion of algal bio-oil produces lower levels of particulates, carbon monoxide, soot, hydrocarbons, and $SO_x$ as compared with petroleum fuels [122].

**Table 7.** Comparison of the properties of moisture-free algal bio-oil, wood bio-oil and petroleum oil

| Property | Bio-oils | | Petroleum |
| --- | --- | --- | --- |
| | Wood | Micro-algae | |
| C | 48.29-56.43 [122,130] | 64.5-82.0 [131] | 83.0-87.0 [122] |
| H | 5.2-7.2 [121] | 7.1-10.4 [131] | 10.0-14.0 [122] |
| O | 30.5-35.3 [130] | 5.3-25.0 [131] | 0.05-1.5 [122] |
| N | 0.31-2.36 [122,130] | 2.5-7.3 [131] | 0.01-0.7 [122] |
| S | 0.01-0.5 [121,130] | 0 [129] | 0.05-6.0 [129] |
| Density [kg/L] | 1.2 [122] | 0.97-1.04 [131] | 0.75-1.0 [122] |
| Water content % | 15-30 [121] | 26.6 [128] | - |
| Viscosity [Pa.s] [122] | 0.04-0.20 (@40°C) | 0.10 (@40°C) | 2-1000 |
| pH | 2.8-3.8 [121] | 4.3 [128] | - |
| HHV [MJ/kg] | 21 [122] | 27.1-29.8 [131] | 42 [122] |

HHV= Higher heating value (Amount of heat produced by the complete combustion of a unit quantity of fuel, producing water, i.e., includes latent heat of evaporation of water).

Algal oil production technology has the potential to yield several billion gallons of renewable fuel per year in the U.S. alone [132,133]. Despite the vast energy potential of micro-algae lipid, data on production costs of algal-derived biofuels are limited to biodiesel, while the current production cost of algal biofuels does not favor the scale-up economics [127,132]. According to a 2010 report by Energy Biosciences Institute (EBI), the capital cost for a 250-acre production system for algal-derived biofuel was about $(US)21 million, with an operating cost of about $(US)1.5 million per year. This could produce about 12,300 barrels of oil, offering a break-even price of $(US)330 per barrel of oil (based on an 8% capital charge). But with 1,000 acres production system of algae biofuel, the break-even price was reduced to about $(US)240 per barrel. Moreover, about 20% of cost reduction was realized when wastewater treatment credits were applied. Although wastewater treatment is considered as the only important co-product for (algal) biofuels production, it is claimed that its availability would significantly reduce the production of algae at the national level [132,133]. Considering the ongoing advances in research and development on biofuels there are potential for technologies for producing fungible, transportation fuel, including gasoline, diesel oil, and jet fuel by using algae [1,132-134]. It was anticipated that Sapphire Energy could soon scale up the process, producing millions of gallons of fuel at prices competitive with fossil fuels. However, major oil price declines in 2008 and subsequently in 2014, made it difficult for biofuels projects to compete with the fossil fuel industry. Other technical problems hampering the development of algal biofuels include maintenance of suitable growing conditions in open ponds, the volumes of pure water, carbon dioxide and fertilizer required to allow the algae to photosynthesize fast enough at large scales and the energy balance of lipid extraction [135].

## 6. Use of abandoned agriculture lands to grow energy crops

Adaptation to marginal lands is one of several benefits of using energy crops (perennial species) for biofuels production, especially with the increasing demand for biomass energy. Energy crops can be cultivated on pasture lands, underutilized cropland, and land currently used for traditional crops. However, using abandoned cropland or pasture lands for energy crops cultivation could be a more viable option than using agriculture lands or the conversion of forested areas [1,66], because these lands require some maintenance to restore their balance of nutrients. Therefore, growing energy crops on marginal lands can restore some of these nutrients by increasing the organic carbon concentration and nitrogen in soil.

## 7. Urban waste as feedstock for biofuels production

Cellulosic municipal solid waste (MSW) and construction and demolition (C&D) wood waste are additional potential feedstock for cellulosic ethanol production. Typically, C&D debris contain 20-30 Vol.% of wood waste, making them potential sources for biofuels production.

## 8. Composition of cellulosic polymeric biomass

Cellulose is the most abundant polymeric, organic matter on earth. The conversion of cellulose polymeric biomass into biofuels may sustainably assist many countries, states, and municipalities toward solving their energy problems. However, the conversion of lignocellulosic biomass into liquid fuels is hindered by the structural and chemical complexity of the biomass. Cellulosic biomass is contained in the matrix of two other natural polymers (hemicellulose and lignin), establishing a complex and heterogeneous material as shown in Figure 15.

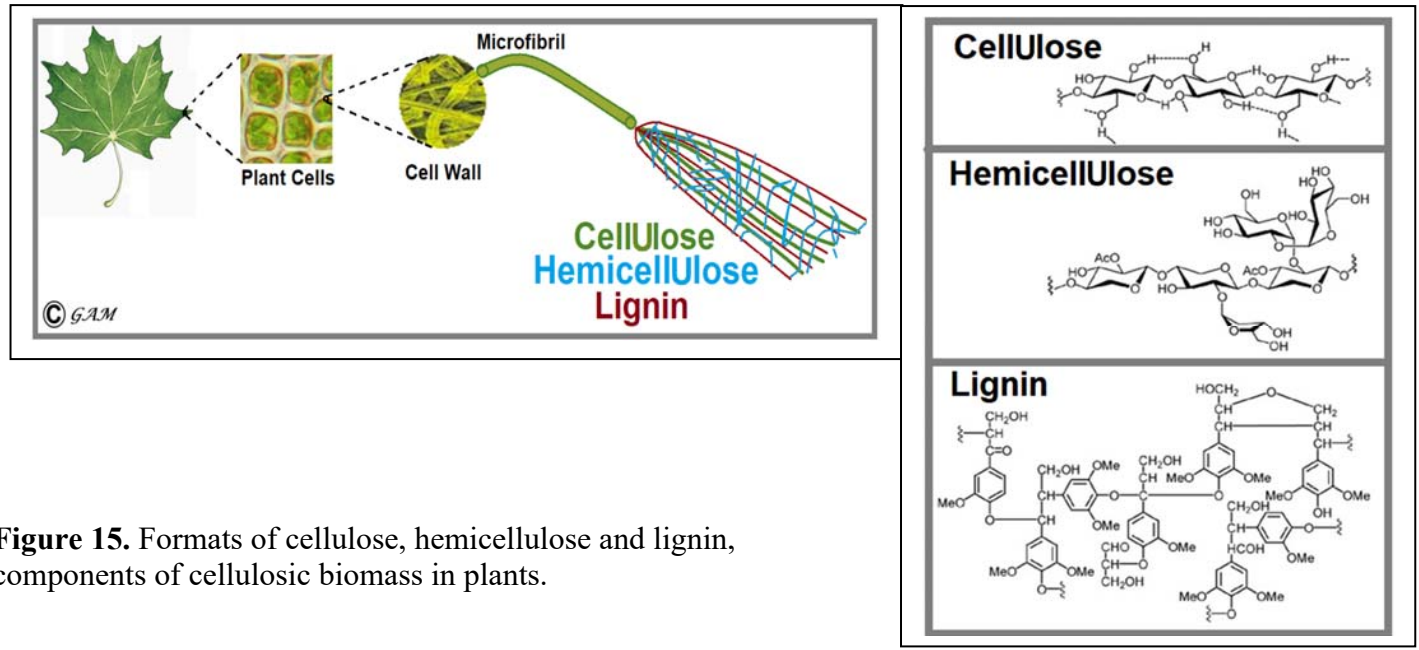

**Figure 15.** Formats of cellulose, hemicellulose and lignin, components of cellulosic biomass in plants.

As shown in Figure 15, cellulose is a molecule of glucose with one molecule of water absent ($C_6H_{10}O_5$). It is the most abundant polysaccharide with ß-glycosidic linkages, but nonetheless, a difficult material to hydrolyze because of its high crystallinity. Hemicellulose, on the other hand, is a relatively weak random and amorphous structure that is easily hydrolyzed using dilute acid or base or cellulase enzymes. Lignin is a relatively strong complex chemical compound occupying the spaces between cellulose and hemicelluloses in the biomass. Lignin is an excellent solid source of fuel with higher energy density than cellulose, but it is not amenable to enzymatic dissociation because of its heterogeneous composition. Table 8 reports the ranges of compositions of cellulose, hemicellulose, and lignin in various cellulosic polymeric biomasses.

**Table 8.** Chemical composition of cellulosic polymeric biomass (on % dry weight basis)

| Biomass Source | Cellulose | Hemicellulose | Lignin | Protein |
|---|---|---|---|---|
| *Crop Residues* | | | | |
| Wheat Straw | 34.76 - 42.08 [138] | 18.13-25.51 [138] | 18.6-23.66 [138] | 5.07 [140] |
| Rice Straw | 37.99-44.61 [138] | 15.03-21.39 [138] | 15.02-21.2 [138] | 4.56 [140] |
| Cotton Stalk | 34.22-42.52 [138] | 9.98 -16.04 [138] | 24.66-30.4 [138] | 5.80-6.90 [141] |
| Corn Stover | 31-38 [79] | 19-44 [79] | 13-21 [79] | 4.05 [144] |
| Corn Fiber Hulls | 15.30 [139] | 40.4 [139] | 2.87 [139] | – |
| Soybeans Hulls | 29-51 [137] | 10 – 25 [137] | 1–4 [137] | 10–15 [136,137] |
| *Other Biomasses* | | | | |
| Wood residues [136] | 40 – 65 [138] | 25 – 40 [138] | 20 -60 [138] | 1.5-2.0 [76] |
| Switchgrass | 44-51 [79] | 42-50 [79] | 13-20 [79] | 3 [85] |
| Miscanthus | 40-60 [84] | 20-40 [84] | 10-30 [84] | 2.5 -5 [80] |
| Industrial Hemp Stalk | 70-74 [142] | 15-20 [142] | 3.5-5.7 [142] | 5.3 [143] |

# 9. Technologies for biofuels production

In this section, the technologies for conversion of biomass into biofuels will be briefly reported. Details of these technologies are presented in [1].

## 9.1. Fischer-Tropsch synthesis (FTS) to producing biofuels

Fischer-Tropsch synthesis represents a mature technology to produce fuels and chemicals. The FTS is carried out using heterogeneous catalytic reactions, usually conducted on supported metals [1,5]. Cobalt (Co), iron (Fe) and ruthenium (Ru) are usual catalysts for FTS, generally supported on silica or alumina. FTS converts synthesis gas (syngas) obtained generally from the gasification process into chemicals through two main reactions during which the original large molecules break up, generating monomers that polymerize, yielding larger hydrocarbon molecules. Syngas obtained from biomass is often referred to as biogas. Synthesis gas is a mixture comprising primarily of hydrogen, carbon monoxide and very frequently smaller amounts of carbon dioxide and short-chain hydrocarbons. Typically, the synthesis gas from a biomass gasifier contains 1-1.5% tar, 0.2-0.4% ammonia, and 100-500 ppm $H_2S$. The selection of suitable operating conditions in the gasifier could reduce tar and ammonia formation. The preferred method for ammonia removal is to decompose it to $N_2$ and $H_2$ using traditional catalysts such as supported Ru, Ni or Fe [1,145].

## 9.2. Fast pyrolysis for biomass conversion to bio-oil

Pyrolysis is a thermochemical process that involves thermal degradation of material at high temperatures and in the absence of oxygen. The pyrolysis process occurs typically at atmospheric pressure and at temperatures ranging from 400 to 800 ºC [1,5]. The products from the pyrolysis process include bio-oil, gas and char. Fast pyrolysis is carried out at a high heating rate, maximizing liquid, bio-oil, products. At short residence times (1-2 seconds) and high heating rates (i.e., 500 ºC/sec), the pyrolysis process produces high content of bio-oils, containing up to 70% of the original energy in the biomass feedstock.

Products from a maximized bio-oil liquid fast pyrolysis reactor could contain about 65 wt% organics, about 10 wt% water and the rest gas and char. However, the relative quantities of these fractions are highly dependent on reactor design, reaction conditions and biomass alkali content. The chemical species in the bio-oil leave the pyrolysis reactor either in the form of vapor, as a free radical precursor, or in the form of aerosol [1,5,145].

## 9.3. Hydrothermal liquefaction (HTL) process for biomass conversion to bio-oil

Hydrothermal liquefaction (HTL) is another thermochemical process for synthesizing of crude bio-oil (biofuel) through the application of heat and pressure on biomass. In this process, the high moisture biomass is subjected to mild temperatures (250-350 ºC) and pressures (10-20 MPa), which break down the high-molecular block structure of the biomass producing bio-crude oil. At these temperatures and pressures, the water in biomass becomes a highly reactive medium, which accelerates the breakdown and cleavage of the chemical bonds. This conversion technology can be used for a broad range of feedstock and it is similar to the naturally occurring geological processes postulated to have led to the formation of underground fossil fuel reserves [2]. Feedstock that is used in this technology include, swine manure, low lipid algae, sawdust, garbage, and even sewage sludge. The products exiting the liquefaction process consist of gas, an aqueous phase containing water-soluble products, and bio-oil. A sample of the product from the liquefaction process relative to the initial biomass feedstock consists typically of 45 wt% bio-oil, 25 wt% gas (with > 90 wt% of gas being $CO_2$) and 30 wt% aqueous phase (with the aqueous phase containing 66 wt% water, and 33 wt% soluble organics) [1,145,146].

## 9.4. Biodegradation process for biomass conversion to ethanol

The biodegradation process combines chemical and biological conversions to produce chemicals from biomass. Pretreatment is required to make the cellulose and hemicelluloses in the biomass more accessible to enzymatic attack for conversion into simple sugars [147,148]. In the traditional biomass biodegradation into ethanol, the pretreatment is generally one of the most expensive processing steps. Recently, new methods have been proposed to reduce the pretreatment cost. One method is by altering the chemical structure of lignin in the biomass crop through genetic engineering. For example, pretreatment enzymes can be produced within the leaves and stems of the biomass crops to lower the amount of lignin as it grows [149]. Furthermore, recent research and development advances have simplified the traditional biodegradation of the biomass process by eliminating and consolidating some of the steps. As a result, there is a direct conversion of cellulose and hemicellulose into ethanol [1,64].

### 9.5. Combined gasification and biodegradation process for ethanol production

Gasification coupled with biodegradation process converts syngas into pure ethanol (or pure butanol) using micro-organisms known as gas fermentation. The combined process enables the lignin in the biomass feedstock, along with hemicellulose and cellulose, to be converted to ethanol. The bacteria or micro-organisms employed for the bioconversion of syngas mixtures to alcohol are referred to as obligate anaerobes, as they require oxygen to grow. The microorganism uses a heterofermentative version of the acetyl-CoA (acetyl group plus Coenzyme A) pathway for acetogenesis. Acetogenesis is a process through which acetate is produced by anaerobic bacteria from a variety of energy and carbon sources. In this pathway, acetyl-CoA is produced from CO or $H_2/CO_2$ mixtures. The acetyl-CoA intermediate is subsequently converted into either ethanol (or acetic acid) as a primary metabolic product [1,150,151].

### 9.6. Renewable liquid fuel production from biomass-derived oils using hydroprocessing

Renewable liquid fuel can be obtained from biomass-derived oil hydroprocessing, as an enhanced method to improve the bio-oil quality. Catalytic hydroprocessing of biomass-derived oils (such as vegetable oils or animal fats) to produce hydrocarbon fuels is a growing trend in the biofuels industry. The fuels resulting from the oil hydroprocessing are, in general, similar in chemical composition and energy content to standard gasoline, diesel, and jet fuel derived from petroleum. The process uses hydrogen and a catalyst (like palladium) to remove oxygen from the triglyceride molecules in the oil feedstock, via decarboxylation and hydrodeoxygenation reactions to produce some light petroleum products and water as co-products [1,66].

### 9.7. Municipal solid waste (MSW) for power generation

Another form of bioenergy that could help communities attain energy-independence from using fossil fuels is the conversion of municipal solid waste (MSW) to biofuels using waste-to-energy conversion plants. Such plants include anaerobic digestion of MSW in the landfill and small-scale digesters, excluding incinerators. The reason for the use of these plants is that from the process engineering point of view, incineration plants are comparable to conventional coal-fired power plants that emit an appreciable amount of pollutants into the environment [1].

### 9.8. Power generation using landfill gas-to-energy technology

The conversion of landfill biomass to energy is a biological conversion process. In the absence of air, anaerobic bacteria in landfills decompose animal manures, organic wastes, and crops to produce biogas. Biogas contains methane, a major component of the natural gas that can be burned to produce heat and electricity. Depending on the origin of the anaerobic digestion process, the composition of biogas can be variable, with a typical landfill gas containing about 40-75% methane by volume. Biogas also contains about 45% $CO_2$ by volume, including a small amount of hydrogen sulfide and ammonia. After purification and upgrading processes, the biogas can be used for various applications similar to natural gas [1,152,153].

### 9.9. The biorefinery of the future and biofuels production industry

A biorefinery is an installation for renewable material utilization that leads to the production of fuels and other bioproducts, similar to the traditional oil refineries (Figure 16). In essence, the term biorefinery refers to a facility or group of facilities combining the production of materials, chemicals, or fuel products with energy production [121]. Biorefineries of the future will upgrade bio-oils using technologies such as filtration, hydrotreating and hydrocracking to reduce the oxygen content in the bio-oil. The final refining processes convert the bio-oil into hydrocarbon fuel products.

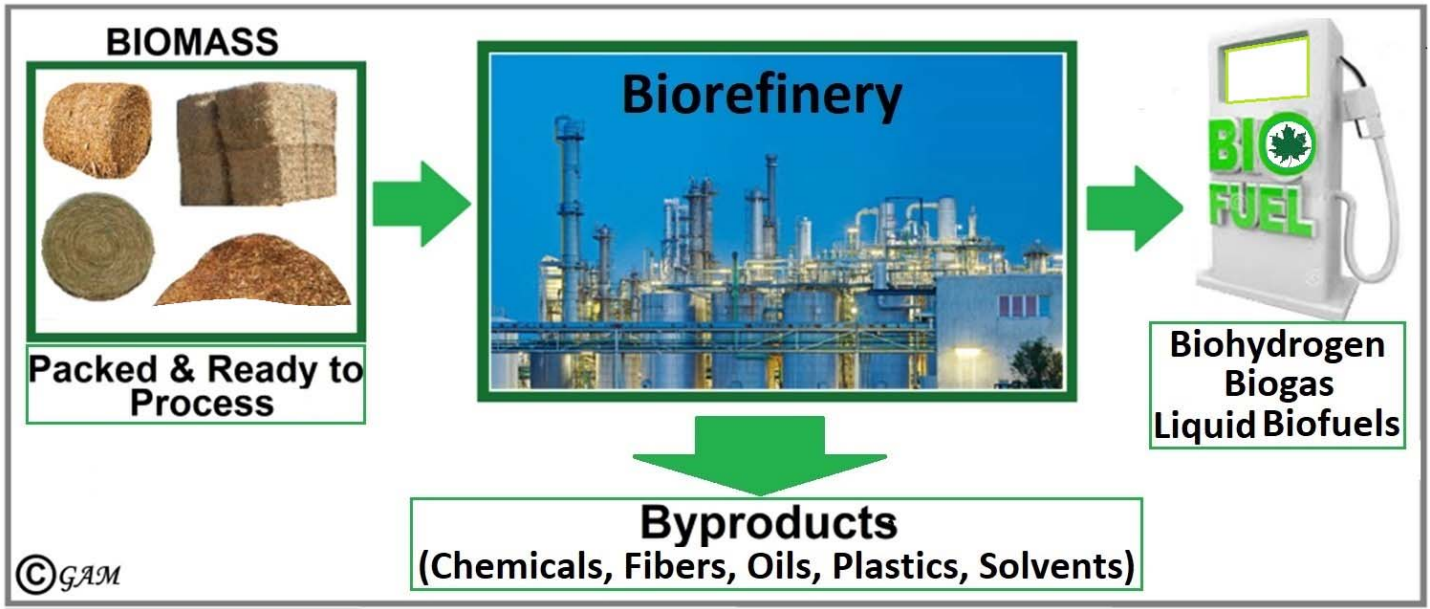

**Figure 16.** Schematic of a biorefinery

Integrated biorefineries are comparable to conventional refineries in that both technologies can produce a broad range of products that can optimize production economics and the use of feedstock. However, integrated biorefineries employ novel technologies and different biomass feedstock that require significant investments in research and development, as well as the deployment of projects to offset the cost involved [1].

## 10. Conclusions

We undertook this project to establish the potential and roles of ammonia, biofuels, and hydrogen, which are considered as non-fossil fuels, for renewable energy sources utilization in the future. Authors of this report have had lifetime experiences all over the world, collectively in all aspects of energy, as well as conventional and modern biological, chemical, and physical engineering and science in the forms of industrial and academic research, including educational activities. That gave us the opportunity to put our experiences together to produce this comprehensive report, which is aimed to help the future generations.

Renewable energies are becoming increasingly important in addressing many of the key challenges facing today's global energy and environmental problems, including prevention of further climate change, cost of energy and energy independence. As it is well documented utilization of renewable energies is increasing across the globe. Ammonia, biofuels, and hydrogen are truly outstanding fuels for large-scale utilization, storage, and transport of renewable energies. Of course, there are major challenges to meet in order to make them of practical everyday use [1,154].

Environmental issues, economics and ease of social acceptance are the three interconnected pillars of sustainability in using ammonia, biofuels, and hydrogen as the fuels of renewable energies, which must be considered for their successful applications. The cost subsidies of government and governmental regulations have been vital to the sustainability of any alternative fuel in the marketplace.

Production of ammonia, biofuels and hydrogen using renewable energies and advancement in their technologies will allow their direct use in the existing internal combustion engines and there will be no need for major electrification of the transportation industry which requires expensive and rare materials, such as lithium, and silver to have high-capacity batteries.

Overall, in a carbon dioxide-constrained world, advanced production, transport and utilization of ammonia, biofuels, and hydrogen: (i). Will contribute to resolving energy shortage. (ii). Will improve the economic viability of renewable energy industry. (iii). Will reduce fossil-fuel-dependence. (iv). Will be beneficial to, both, the environment, and the society in terms of reduction of greenhouse gas emissions and global warming. (iv). Will serve the green transportation industry, and (v). Will provide a competitive advantage to the renewable energy industry.

**Acknowledgements:** The data reported in the tables of this paper were collected by L. Barnie Agyarko.

26. Kobayashi, H.; Hayakawa, A.; Somarathne, K.D.K.A.; Okafor, E.C. Science and technology of ammonia combustion. *Proceed. Combustion Inst.* **2019**, 37, 109–133.

27. Okafora, E.C.; Somarathne, K.D.K.A.; Hayakawa, A.; et al. Towards the development of an efficient low-NOx ammonia combustor for a micro gas turbine. *Proceed. Combustion Inst*. **2019**, 37, 4597-4606.

28. The Engineering ToolBox. **1987**, Heat of combustion. (www.engineeringtoolbox.com/standard-heat-of-combustion-energy-content-d_1987.html)

29. The Engineering ToolBox. **2020**, Liquid Densities: Densities of common liquids like acetone, beer, oil, water and more. (www.engineeringtoolbox.com/liquids-densities-d_743.html)

30. College of the Desert. *Hydrogen Properties. Hydrogen Fuel Cell Engines and Related Technologies* **2001**.

31. TRS. Ammonia: zero-carbon fertilizer, fuel and energy store: Policy briefing. *The Royal Society* **2020**, 40 pages. ISBN: 978-1-78252-448-9.

32. Brown, T. Ammonia turbines, industrial furnaces, fuel-cells. *Ammonia Energy Assoc.* **2017**.

33. Lewis, J. Fuels Without Carbon. *Clean Air Task Force* **2018**, Boston, MA, USA.

34. Gladyshev, G. On the thermodynamic direction of the origin of life and its evolution: a new confirmation of the theory. *Norwegian J. Devel., Int'l Sci.* **2019**, 2, 31-36.

35. Pattabathula, V.; Richardson, J. Sixty Years of History of the AIChE Ammonia Safety Symposium. *Proceed. 60th Ann. Safety in Ammonia Plants & Related Facilities Symp*. **2015**, pp. 1-26, AIChE, New York.

36. Tiseo, I. Biofuels - production worldwide 2000-2019. *Statistica,* July 23, **2020**.

37. Lorne, D.; Bouter, A. Economic Outlook: *Biofuels Dashboard* **2019**, IFP Energies Nouvelles, France.

38. Probstein, R.F.; HICKS, R.E. *Synthetic Fuels*. **1982**, McGraw-Hill.

39. US EPA, U.S. Energy Independence and Security Act of 2007. US Env. Prot. Agency **2007**.

40. Moravvej, Z.; Makarem, M.A.; Rahimpour. M.R. The fourth generation of biofuel. In book "*Second and Third Generation of Feedstocks*" **2019**, pages 557-597, ISBN 978-0-12-815162-4. Elsevier.

41. Wang, J.L.; Mansoori, G.A. A revision of the distillation theory. *Scientia Ir*. **1994**, 1, 267-287.

42. Mansoori, G.A. Fundamental thermodynamics for distillation Proc. *Proceed. Distillation Seminar*, *Lect. 1,* **1999**, AIChE-Chicago Sect., Chicago, IL, 22 pages.

43. USDA, A USDA regional roadmap to meeting the biofuels goals of the renewable fuels standard by 2022. *U.S. Dept. of Agric*., Washington, DC. **2010**, 21 pages.

44. Unnasch. S. GHG Reductions from the RFS2: A 2018 Update. *Life Cycle Associates* **2019**, Report No. LCA.6145.199.2019. Prepared for Renewable Fuels Assoc.
32